# A Knowledge Graph and a Tripartite Evaluation Framework Make Retrieval-Augmented Generation Scalable and Transparent

**Olalekan K. Akindele, Bhupesh Kumar Mishra\*, and Kenneth Y. Wertheim**
(o.k.akindele-2022@hull.ac.uk, bhupesh.mishra@hull.ac.uk\*, k.y.wertheim@hull.ac.uk)
Centre of Excellence for Data Science, Artificial Intelligence, and Modelling (DAIM), and the School of Computer Science; University of Hull, Cottingham Road, HU6 7RXHull, United Kingdom.
\*Corresponding Author: Bhupesh Kumar Mishra/ Kenneth Y. Wertheim,
email: bhupesh.mishra@hull.ac.uk/K.Y.Wertheim@hull.ac.uk

**Abstract**

Large Language Models (LLMs) have significantly enhanced conversational Artificial Intelligence (AI) chatbots; however, domain-specific accuracy and the avoidance of factual inconsistencies remain pressing challenges, particularly for large datasets. Designing an effective chatbot with appropriate methods and evaluating its effectiveness is among the challenges in this domain. This study presents a Retrieval-Augmented Generation (RAG) chatbot that harnesses a knowledge graph and vector search retrieval to deliver precise, context-rich responses in an exemplary use case from over a hundred thousand engineering project-related emails, thereby minimising the need for document chunking. A central innovation of this work is the introduction of RAG Evaluation (RAG-Eval), a novel chain-of-thought LLM-based tripartite evaluation framework specifically developed to assess RAG applications. This tripartite framework operates in parallel with the chatbot, jointly assessing the user's query, the retrieved document, and the generated response, enabling a holistic evaluation across multiple quality metrics—query relevance, factual accuracy, and coverage, as well as coherence and fluency. The resulting scoring system is provided directly to users as a confidence score (1 – 100%), enabling quick identification of possible misaligned or incomplete answers. This proposed approach promotes transparency and rapid verification by incorporating metadata (email IDs, timestamps) into responses. Experimental comparisons against BERTScore and G-EVAL for summarisation evaluation tasks confirm its effectiveness, and empirical analysis also shows RAG-Eval reliably detects factual gaps and query mismatches, thereby fostering trust in high-demand, data-centric environments. These findings highlight a scalable path for developing accurate, user-verifiable chatbots that bridge the gap between high-level conversational fluency and factual accuracy.

## 1. Introduction

Recent advancements in Artificial Intelligence (AI), particularly large language models (LLMs) such as OpenAI's GPT [1], Meta's Llama [2], and Google's Gemini [3], have dramatically expanded the capabilities of conversational AI across diverse domains. Their extensive pre-training on vast text corpora enables them to produce coherent, context-aware responses without task-specific fine-tuning. However, this broad coverage can inadvertently cause domain inaccuracies in specialised areas requiring precise contextual understanding. To mitigate such issues, Retrieval-Augmented Generation (RAG) [4–8] has emerged as a powerful technique that integrates external retrieval mechanisms with generative models, grounding LLM outputs in relevant source materials rather than relying solely on the model's internal parameters and training data.

Despite the promise of RAG for improving information retrieval and domain-specific accuracy, practical deployments still face two major obstacles. First, reliably retrieving the correct information from large text documents can be challenging; conventional RAG approaches often break documents into smaller chunks, risking loss of contextual integrity in domains where relationships between data points are crucial. Second, evaluating RAG applications, such as chatbot responses in real time remains difficult. Traditional Natural Language Processing (NLP) metrics such as Recall-Oriented Understudy for Gisting Evaluation (ROUGE) [9] and Bilingual Evaluation Understudy (BLEU) [10] primarily focus on token-level overlap. In contrast, embedding-based metrics such as BERTScore [11] use contextual representations to assess semantic similarity between responses. However, both classes of metrics have limitations for end-to-end chatbot evaluation. Recent LLM-based metrics, such as LLM-retEval [12], focus on retriever evaluation for Question Answering (QA) tasks but do not assess overall query alignment or detailed factual coverage.

To address these gaps, this work presents the results of an advanced RAG project [13–17], a RAG chatbot combining a knowledge graph with a vector database in the retrieval step, and a new real-time tripartite evaluation framework called RAG-Evaluation (RAG-Eval). During operation, the framework receives the user's query, the retrieved document, and the chatbot's generated response as inputs. It evaluates these components together to provide the user with a transparent and contextually grounded assessment of each response. Together, these components provide more accurate, trustworthy, and transparent conversations about large and structured datasets. Specifically:

(i) **Knowledge Graph and Vector Search Retrieval:** Developed RAG chatbot system underpinned by a knowledge graph and vector-based retrieval, designed to retrieve information from a hundred thousand project-related emails in a context-sensitive manner. To harness semantic relationships within the corpus, the approach avoids heavy reliance on document chunking and preserves fine-grained details essential to correct question answering.
(ii) **RAG-Eval for Real-Time Response Assessment**: Developed a tripartite RAG-Evaluation (RAG-Eval) framework, an LLM evaluation Python package that analyses responses based on five key metrics—query relevance, factual accuracy, coverage, coherence, and fluency. All metrics are assessed jointly by comparing the user's query, the retrieved data, and the generated response. This integrated perspective helps detect misalignments and inconsistencies often overlooked by standard evaluation metrics.
(iii) **Metadata-Enriched Summaries for Transparency**: To promote explainability and user trust, each chatbot response is accompanied by metadata (email IDs, timestamps, etc.) that enables immediate cross-referencing against the source. This design element addresses a common shortcoming of many RAG systems, where users lack transparent evidence for verifying chatbot outputs.

Evaluation results support the claim that the proposed system can leverage large-scale structured datasets to generate domain-specific answers to unseen questions and is also comprehensively capable of assessing response quality in real-time. These features collectively make the RAG chatbot scalable, reliable, and suitable for high-demand settings where both precision and transparency are paramount.

The rest of the paper is organised as follows: Section 2 provides a comprehensive review of related works; Section 3 shows how information flows within the RAG chatbot system, including RAG-Eval; Section 4 details data preprocessing, knowledge graph construction, embedding generation, a detailed introduction to RAG-Eval, and metadata integration techniques; Section 5 presents the results of evaluating the chatbot responses using various queries, comparing RAG-Eval with established metrics in the context of summarisation, and evaluating query mismatches and factual alignment; Section 6 explains the key contributions and limitations of the RAG chatbot system and RAG-Eval; and Section 7 concludes the paper with the future avenues of research.

**2. Literature Review**

The evolution of chatbot technology has been propelled by continuous breakthroughs in AI and NLP [18–20]. Early systems often relied on dialogue flow-based or rule-based designs, which were adequate for simple queries but floundered under more complex conversational demands [21,22]. As user expectations grew, these methods were limited by their rigid structures and inability to handle complex, context-rich queries. The advent of LLMs such as GPT, Llama, and Gemini has dramatically expanded the scope of automated dialogue, enabling more coherent and context-aware interactions [1–3]. Nonetheless, even these powerful models are not without shortcomings, often exhibiting issues like factual inaccuracies (known as hallucination [23]) or responding to queries that involve domain knowledge. Various studies have been carried out to investigate when/why factual gaps emerge and how they can be mitigated. For example, benchmarks like HaluEval 2.0 [24] can be used to track factual inaccuracies across different architectures, revealing how independent factors such as training data and prompts affect LLM outputs. Others have employed entropy-based metrics to statistically pinpoint deviations from verifiable truths [25]. While including such techniques improves reliability, scaling LLMs to real-world applications presents their own challenges, including memory constraints [26] and the diminishing returns of mere parameter growth [27]. Domain-specific fine-tuning has also proven tricky, so alternative strategies are necessary. For example, frameworks such as DragFT [28] exemplify how carefully selected prompts and retrieval-based techniques can improve a model's performance without overspecialising it, and comprehensive data–curation processes can help a model respond to queries within niche domains [29]. These techniques are components of a general strategy called Retrieval-Augmented Generation (RAG).

RAG has emerged as a promising solution to these issues by grounding LLM outputs in relevant and real-time knowledge sources. For example, in finance, query expansion and metadata annotations can potentially result in more precise document analysis [30]. Similarly, healthcare applications could benefit from RAG. In one study, OpenAI embeddings were used to guide patient-facing explanations for surgical procedures, ultimately improving comprehension [31]. RAG has found success in data-intensive scenarios, including the multi-LLM system ERATTA for enterprise environments [32]. However, RAG typically relies on chunking documents, risking loss of relational integrity and reduction of retrieval accuracy. Knowledge graphs (KGs) present a more structured alternative because they retain explicit relationships among data points. Some researchers have integrated KGs into RAG frameworks to significantly improve the precision and depth of retrieval [33]. Others [34] propose hybrid solutions that fuse structured and unstructured data for iterative retrieval, enabling large language models to engage in deeper reasoning steps. Such design choices help mitigate data fragmentation, thereby minimising inaccurate or incomplete responses in chatbot dialogues.

Despite these advances in retrieval, evaluations of RAG applications, such as chatbot responses often fail to capture essential QA dimensions such as query alignment, provenance, and fine-grained fact inclusion. Traditional metrics such as BLEU and ROUGE measure only token overlap [9,10]. Embedding-based scores such as BERTScore [11] compare candidate text to a reference summary without checking whether a response is relevant to the user's question. G-EVAL [35] leverages GPT-4's chain-of-thought prompting to score candidate summaries or responses against their source documents (summarisation or dialogue). LLM-retEval [12], on the other hand, evaluates the retriever's contribution by comparing answers generated with retrieved passages to those generated from gold-labelled passages—passages that have been manually identified as containing the correct information for answering the query. However, like other methods, its evaluation scope is limited; it does not explicitly assess answer coherence, fluency, or comprehensive coverage of all query-relevant facts. In short, none of these evaluation metrics is suitable for evaluating the full "(user query + retrieved context + generated response)" tuple along three important dimensions: query relevance, factual accuracy, and coverage of the essential details. RAG-Eval fills this gap by providing a unified, tripartite evaluation framework that evaluates chatbot performance using all three key components of the RAG process: the user query, the retrieved context, and the generated response. This joint assessment across multiple quality metrics—query relevance, factual accuracy, coverage, coherence, and fluency.

## 3. System Architecture and Workflow Overview

Fig. 1 presents the end-to-end architecture of the proposed RAG chatbot system. The workflow begins with user queries submitted via a React-based chatbot interface, routed to a FastAPI backend that orchestrates communication with a Langchain ReAct AI Agent. The AI Agent, powered by an LLM, retrieves information from the Neo4j graph database via a primary Cypher query or, if needed, vector-based semantic search. Based on the retrieved information, the LLM formulates a response, which is then evaluated by RAG-Eval, which operationalises a tripartite evaluation framework by computing a confidence score (overall accuracy score) based on principal criteria—query relevance, factual accuracy, coverage, coherence, and fluency. The final response, along with the confidence score, is returned to the user interface via the FastAPI backend, ensuring transparent and reliable interactions. The following subsections provide a brief overview of each architectural component and describe the operational workflow that enables accurate, transparent, and efficient chatbot interactions.

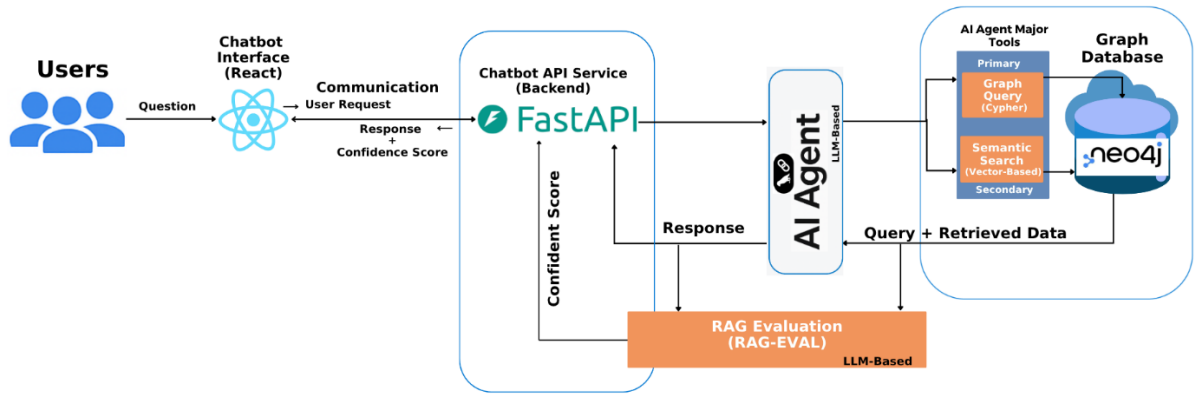

**Fig. 1** Chatbot interaction and response generation flowchart

**3.1 Overview of the RAG Chatbot Architecture**

This study employs the RAG architecture that integrates a knowledge graph with a vector database to enhance search retrieval and thus produce accurate responses about engineering project communications. It is also equipped with the RAG-Eval framework in which its responses are evaluated. The key components of this RAG chatbot architecture include:

- **Knowledge Graph (Neo4j):** Structures project data into interconnected nodes and relationships, enabling the chatbot to respond to complex and context-dependent queries. Facilitates efficient data retrieval and enhances the chatbot's ability to understand and navigate intricate project information.
- **Vector Search Retrieval:** Utilises semantic embeddings stored in the graph database to perform similarity searches (cosine similarity search), retrieving information relevant to user queries.
- **AI Agent:** Processes the queries, generates Cyphers for the retrieval, and formulates responses based on the retrieved context.
- **RAG Evaluation (RAG-Eval):** To support transparent and reliable assessment of chatbot responses, a dedicated LLM-based tripartite evaluation framework, RAG-Eval, was developed. Operating independently of the chatbot's response generation process, RAG-Eval receives the user query, the retrieved document, and the system's generated response as inputs. It then evaluates these jointly, providing the user with a comprehensive, multi-metric confidence score. This process ensures that evaluation is grounded in all available context, allowing users to gauge the quality and reliability of each response immediately.
- **Frontend (React) and Backend (FastAPI):** The React frontend provides an interactive and dynamic user interface. It communicates with a FastAPI backend via RESTful APIs (Representational State Transfer Application Programming Interface), a common approach where the frontend asks for data or submits actions, and the backend responds. FastAPI manages all interactions between the frontend and the underlying data processing services, including LLM processing, knowledge graph queries, and response evaluation, thereby offering enhanced scalability, modularity, and performance.

## 3.2 Operational Workflow

The operational workflow of the RAG chatbot follows a structured multi-stage process that ensures accurate, contextually relevant responses, supported by real-time quality evaluation through RAG-Eval. The detailed steps involved are as follows:

1. **User Query Processing**: A user types a question into the React-based interface, which sends that query to the FastAPI backend via a Hypertext Transfer Protocol (HTTP) POST request (behind the scenes using a standard web request). FastAPI then forwards the query to the AI agent for processing.
2. **Cypher Query Generation**: Based on the interpreted query, the AI agents generate a Cypher query to retrieve specific nodes and relationships from the graph database. Cypher, Neo4j's native query language, enables precise retrieval of relevant emails and associated metadata such as email IDs, senders, and timestamps.
3. **Vector Search (Secondary)**: If the Cypher query returns insufficient results or fails entirely, the system automatically triggers vector similarity search. It converts the query into a semantic embedding and retrieves similar emails from the Neo4j database.
4. **Data Retrieval**: Retrieved data from either the Cypher query or vector search provides the necessary context, ensuring responses remain grounded in verified sources. The retrieved nodes (email and its properties, persons (sender and recipients), and conversation) are compiled into a structured context and appended to the LLM-based AI agent prompt.
5. **Response Generation and Evaluation**: After retrieval, the LLM-based AI agent uses the ReAct framework to reason over the retrieved context and dynamically decides the optimal response strategy to generate contextually accurate and metadata-enriched answers. The generated response is then evaluated by RAG-Eval, computing the overall accuracy (confidence score) based on the principal metrics—query relevance, factual accuracy, coverage, coherence, and fluency.

**Practical Example Workflow**:

To illustrate, consider a query, *"Who sent the email regarding 'pull tester training' on January 5th, 2022?"* The system generates a Cypher query to retrieve the corresponding documents (Email node and its properties) from the Neo4j graph database:

```
MATCH (sender:Person)-[:SENT]->(e:Email)
WHERE toLower(e.subject) CONTAINS toLower('pull tester training')
 AND e.timeReceived CONTAINS '2022-01-05'
RETURN sender.personId AS senderId, e.emailId, e.revisionId, e.timeReceived, e.content
```

Based on the Cypher query, if no direct match is found, the system falls back to a vector search utilising cosine similarity. The final response is then composed by the AI agent, verified and scored by the RAG-Eval framework, typically returning a confidence score (for example, 100%).

## 4. Experimental Setup and Evaluation Framework

This section details the experimental setup employed to develop and evaluate the RAG chatbot, including data preparation, knowledge graph construction, embedding integration, query generation methods, and the design of the evaluation framework (RAG-Eval), alongside metadata integration for enhanced response transparency.

### 4.1 Data Preparation and Knowledge Graph Construction

### 4.1.1 Dataset Description

The dataset, after preprocessing, comprises over a hundred thousand engineering project-related emails and document revisions from a UK-based company primarily working in the civil engineering sector. The dataset includes rich metadata such as email identifiers (Email IDs), subjects, timestamps, and the full textual content of the emails. This extensive collection provides a solid foundation for constructing a knowledge graph that captures large-scale engineering projects' complex communications and relationships.

### 4.1.2 Data Wrangling and Preprocessing

To prepare the dataset for integration into the knowledge graph and subsequent processing by the chatbot, the following steps have been implemented:

- **Consolidation:** Five structured CSV files containing various aspects of project communications and document revisions were used. The consolidation was performed based on shared keys such as Email IDs and Revision IDs, resulting in a unified dataset that preserves the relational integrity of the original data sources.

- **Parsing Sender and Recipient Names:** Sender and recipient names were not explicitly available in the initial dataset; rather, this information was often embedded within email addresses or found in the email content. To extract this information and ensure it is anonymised, email addresses were parsed, and natural language processing techniques and regular expressions (Regex) were leveraged to infer email addresses and names from email content, where possible to isolate personal names and email addresses. This step was crucial for constructing the social dynamics within the knowledge graph, enabling queries about communications between specific individuals or teams. To understand the nature of these exchanges, the emails have been analysed with various visualisations, including a histogram as shown in Fig. 2. The analysis shows that most emails are concise, typically under 500 words, with occasional in-depth messages of over 800 words. This distribution suggests a mix of quick updates and detailed discussions.

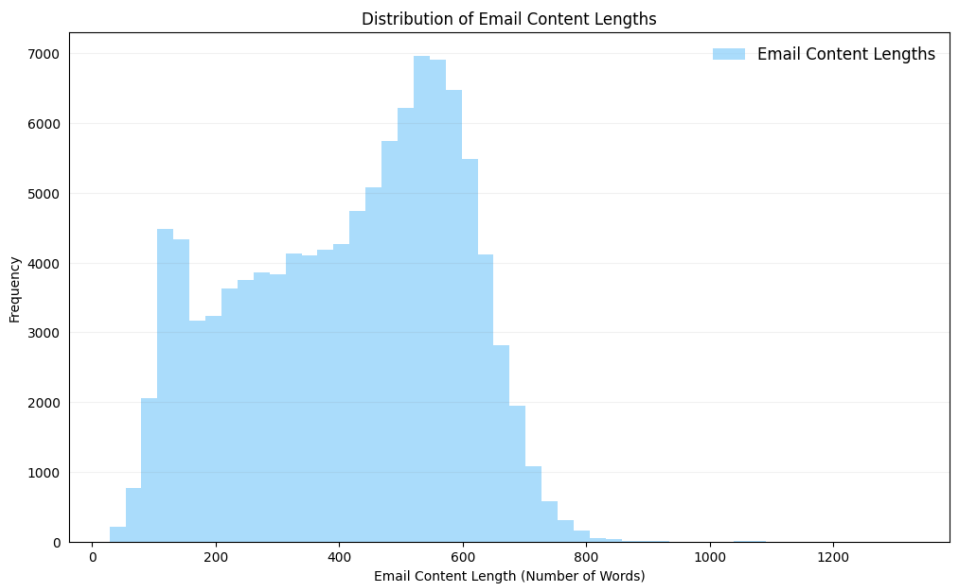

**Fig. 2** Histogram showing the distribution of email content lengths

### 4.1.3   Anonymisation and Pseudonymisation

Considering the sensitive nature of the data, rigorous anonymisation and pseudonymisation [36,37] protocols have been applied to protect confidentiality. For instance, while all email addresses were removed, personal identifiers (sender and receiver names) were irreversibly hashed before being ingested into the graph. Each raw name was concatenated with a secret salt—a randomly generated string and hashed using the SHA-256 algorithm [38–41], producing a one-way, 256-bit digest. It was then truncated to the first twelve hexadecimal characters in each digest to create a deterministic, non-reversible pseudonym. The final dataset retains only these hashed IDs; no plaintext names remain.

In addition, several complementary preprocessing steps were performed to further enhance privacy protection and data cleanliness. These include systematic removal of email signatures and footers, as well as thorough scrubbing of residual personal identifiers within the email content

- **Signature and Footer Removal:**
  To further remove personal details and irrelevant information, all email bodies underwent an additional cleaning step using the Talon package [42,43], a widely adopted tool for detecting and removing email signatures. Talon's hybrid approach, combining pattern recognition and statistical modelling, reliably strips most signature blocks. For any signatures not automatically detected, supplementary regex heuristics targeted common closing phrases and formats. This ensured that the final dataset remained free from signature noise, supporting both privacy and data quality.
- **Personal Identifier Scrubbing and Email Content Cleanup:** To guard against any residual personal data, every email body was processed through a custom cleaning pipeline. Microsoft Presidio's AnalyzerEngine and AnonymizerEngine [44], which detect phone numbers, personal names, email addresses, locations, and postcodes, were used to strip or replace all detected entities with corresponding generic terms.

The data wrangling process resulted in all Personally Identifiable Information (PII) being dropped (email address, sender name, receiver name) and a clean dataset, limited to Conversation_ID, Revision_ID, Email_ID, Document_ID, Sender_ID, Receiver_IDs, Email_Subject, Email_Content, Time_Received, and Time_Modified, ready for embedding generation and knowledge graph construction

### 4.1.4 Embedding Generation and Integration

To facilitate semantic search and improve the RAG chatbot's ability to retrieve relevant information, embeddings have been generated for each email's content. The steps involved are as follows:

- **Embedding Generation:** OpenAI's embedding model (text-embedding-3-small) has been implemented to convert each email's combined text into a high-dimensional vector representation. The combined or concatenated text for each email included the subject, sender, and recipient IDs, timestamps, revision IDs, email IDs, and the email body. This holistic representation ensures that the embeddings capture both the semantic content and contextual metadata of the emails.

- **Integration into the Knowledge Graph:** The embeddings were stored as properties within the corresponding email nodes in the Neo4j graph database. This integration allows for efficient retrieval of emails based on semantic search, enhancing the chatbot's ability to understand and respond to user queries accurately.

- **Vector Index Creation:** A vector index was created on the embedding property using Neo4j's native indexing capabilities. This index facilitates rapid vector similarity searches (cosine similarity) within the graph database, enabling the system to handle real-time queries effectively.

The knowledge graph schema was designed to model the complex relationships present in the dataset. It includes:

- **Nodes:**
    - **Email:** Represents individual emails with properties for content, metadata, and embeddings.

- **Person:** Represents individual persons involved in the communications, serving as senders or recipients.
- **Conversation:** Group emails that are part of the same thread (same conversation ID).
- **Relationships:**
  - **SENT:** Links a Person node to an Email node, indicating the sender of the email.
  - **RECEIVED:** Links a Person node to an Email node, representing the recipient of the email.
  - **PART_OF:** Associates an Email node with a Conversation node, grouping related communications.

By structuring the data in this manner, both the semantic content and the relational context were preserved, enabling complex queries that leverage the strengths of both graph databases and vector-based semantic search. It is also worth noting that the knowledge graph is built and maintained statically before user queries, and all emails are ingested into the Neo4j database. When new data arrives (additional project emails), it can be merged incrementally into the same graph schema, keeping it up to date. This ensures that each user query operates on a stable, pre-existing graph rather than creating or restructuring the graph for every request.

Fig. 3 exemplifies the structure and richness of the constructed knowledge graph, using a subgraph produced by a Cypher query that links Person nodes to Email nodes via SENT and RECEIVED relationships. In this visualisation, olive circles represent Person nodes and the associated Conversation node (orange circle), while light blue circles denote individual Email nodes. The central sender stands out as the Person node connected to multiple sent emails, while the graph also demonstrates how a single email can have multiple recipients (via RECEIVED relationships) and can belong to a particular conversation thread (PART_OF). This slice of the graph captures both the communication networks, who communicate with whom, and the conversational context, how messages are grouped into threads, enabling rich, multi-dimensional querying for the RAG applications.

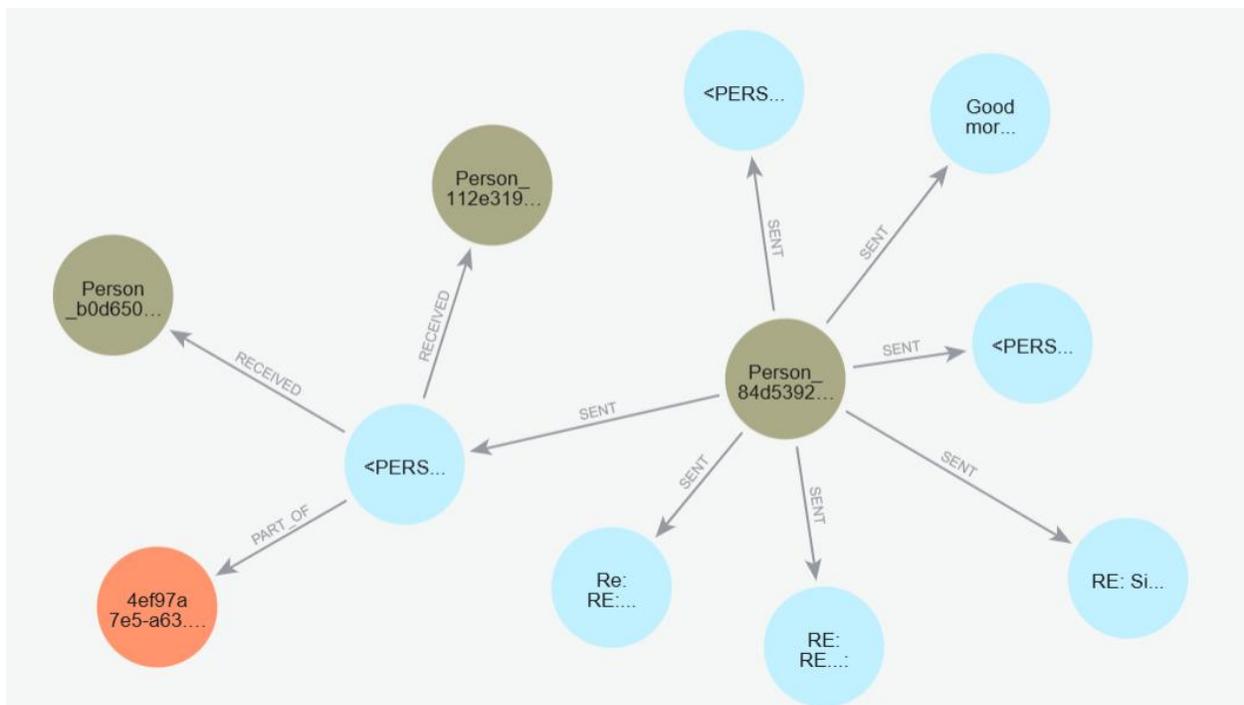

**Fig. 3** Knowledge graph visualisation from Neo4j database console showing anonymised persons (olive), emails (light blue), and the conversation node (orange)

## 4.2 Query Generation, Retrieval, and Response Methods

The chatbot employed a structured multi-stage approach combining Cypher query generation and vector-based retrieval, followed by agent-driven response generation.

### 4.2.1 Cypher Query Generation

Queries were initially handled by generating Cypher queries using the AI agent instructed through carefully engineered prompts. The following system prompts guided this process:

**Cypher Generation Template:**

*You are an expert Neo4j Developer tasked with translating user questions into Cypher queries to retrieve relevant information from an email database.*

***Guidelines:***

1. *Convert the user's question based on the provided schema (Emails, Person, relationships).*
2. *Use only the specified relationship types and properties.*
3. *Do not return entire nodes or embeddings.*
4. *Always return metadata like 'emailId' and 'timeReceived' properties alongside necessary details.*

This approach leveraged schema-specific prompts, ensuring that generated queries were precise, relevant, and adhered to defined database constraints.

### 4.2.2 Vector-Based Retrieval (Secondary)

If the initial Cypher query did not yield sufficient information to answer the user's query, the AI Agent then triggered a secondary vector-based retrieval. This fallback mechanism was prompted only after the agent determined that the primary retrieval did not provide an adequate output or basis for response. The results from this vector search were then used as context for the AI Agent to formulate a response, guided by the instruction:

*"Use the given context to answer the question. If you don't know the answer, say you don't know."*

This ensured the chatbot maintained reliability, delivering contextually relevant information even when direct queries failed.

### 4.2.3 Agent-Driven Response Generation

A ReAct-style agent [45,46] orchestrated the entire retrieval process by first attempting to answer queries using Cypher-based graph search. If the result was insufficient, the agent automatically triggered vector-based search as a secondary strategy. This iterative resolution continued, integrating multiple retrieval methods until a contextually relevant response could be generated or all options were exhausted. If no suitable answer was found, the agent defaulted to a general fallback. This adaptive, agent-driven workflow enabled the chatbot to address queries of varying complexity and clarity. The agent's prompts included:

*"You are a database expert providing information about the email database. Be helpful, provide maximum information, and always include relevant metadata like IDs and dates. Do not answer unrelated queries."*

During the development, zero-shot, few-shot [1,47], and chain-of-thought [48] prompting were also experimented with based on the complexity cues of the queries:

- **Zero-Shot Prompting:** At the start, zero-shot prompting was experimented with, as the model was provided with only a high-level system directive, for instance, "***You are a database expert, generate the best possible***

***Cypher query.***" No examples or additional guidance were included. This approach allowed testing the base capability of the LLM-powered AI agent in interpreting user requests about the email dataset. However, observing that zero-shot prompts sometimes yielded incomplete or slightly off-target Cypher statements, especially for queries involving multiple constraints (like date plus specific sender or subject).

- **Few-Shot Prompting:** To address the limitations of zero-shot, few-shot prompting was employed by including a small set of examples within the chatbot system prompt. For example, implemented multiple successful Cypher queries illustrating typical project-related requests, one focusing on retrieving emails for a certain email ID, another combining sender and topic constraints, etc. These examples gave the agent a clearer template to follow, increasing both accuracy and consistency in the generated Cypher.

Examples of the few-shot prompting implemented for Cypher statements include:

*1. To find the email person where the subject or email content is specified:*

MATCH (sender:Person)-[:SENT]->(e:Email)

WHERE toLower(e.subject) CONTAINS toLower('pull tester training')

   OR toLower(e.content) CONTAINS toLower('pull tester training')

RETURN sender.personId AS senderId, receiver.personId AS receiverIds, e.emailId, e.revisionId, e.timeReceived, e.content

*2. To find emails based on a person's ID and a specified subject:*

MATCH (sender:Person)-[:SENT]->(e:Email)

OPTIONAL MATCH (e)<-[:RECEIVED]-(receiver:Person)

WHERE sender.personId = 'Person_700b29d2c283'

   AND (toLower(e.subject) CONTAINS toLower('Land Contract')

   OR toLower(e.content) CONTAINS toLower('Land Contract'))

RETURN sender.personId AS senderId, receiver.personId AS receiverIds, e.emailId, e.revisionId, e.timeReceived, e.content

*3. To find who sent an email based on the email ID:*

MATCH (sender:Person)-[:SENT]->(e:Email)

WHERE e.emailId = 123456789

RETURN sender.personId AS senderId, e.revisionId, e.timeReceived, e.content

**Chain-of-Thought Reasoning:** For user queries requiring stepwise reasoning over the knowledge graph (such as "Find all emails related to 'pull tester training' and summarise their content"), we experimented with chain-of-thought prompts [48]. Here, the model is explicitly allowed to reason step-by-step, effectively "thinking out loud" before finalising a Cypher query or summarising retrieved data.

- **Implementation Detail**: The LangChain framework [18] implementation of a "react-style agent" approach was used, where the agent's hidden scratchpad (an internal log where the agent writes out its step-by-step reasoning and actions as it works through the query) is not shown to the end-user (only visible on the backend). This reveals how the agent systematically arrives at partial matches ("Check if a conversation node links Email A to Email B," etc.). This chain-of-thought technique increased the correctness of complex queries but also required careful filtering so that verbose reasoning would not clutter the final user-facing output.

**ReAct (Reasoning and Acting) Prompt Structure: Thought, Action, and Rules**

***AGENT PROMPT:***
*You are a database expert providing information about an email database.*

*Be as helpful as possible and return as much information as possible.*

*Do not answer any questions that do not relate to the email database, sender, or receiver.*

*Always include relevant metadata like IDs and dates provided in the context for reference. This helps boost user confidence in your response.*

*Do not output 'I do not know' or similar phrases if you have not used all available tools.*

*Do not answer any questions using your pre-trained knowledge; only use the information provided in the context.*

***TOOLS:***

*You have access to the following tools:*

*=> Email database information (Cypher tool - Translate the question into Cypher and run it against the email schema)*

*=> Email Database Search (Vector search tool - Semantic search over concatenated email content and metadata)*

*=> General Chat (Fallback chat when no other tool applies, for example, a user with a query 'Hello' or 'Hi')*

*To use a tool, please use the following format:*

***THOUGHT AND ACTION:***

*Thought: Do I need to use a tool? Yes*

*Action: The action to take should be one of the tools*

*Action Input: the input to the action*

*Observation: the result of the action*

***RULES:***

*Tool usage rules:*

*1. Always try to use "Email database information" on the cypher query first.*

*2. If "Email database information" returns no context or is not relevant after one try, then use "Email Database Search".*

*3. Use "General Chat" only if the other tools are not suitable for the query.*

*When you have a response to say to the Human, or if you do not need to use a tool, you must use the format:*

*Thought: Do I need to use a tool? No*

*Final Answer: [your response here]*

**4.2.4 Arriving at the Final Prompt Designs**

This work ultimately iterated on these strategies, zero-shot, few-shot, and chain-of-thought, by running the system on a range of real user queries. Each iteration was guided by practically measuring how often the generated Cypher matched the expected results or how closely the summary aligned with the knowledge graph data. After several refinement cycles, we settled on concise few-shot prompts that performed well for most queries, while still allowing chain-of-thought reasoning for more complex cases. In practice, the system can dynamically select a stepwise (chain-of-thought) approach for input queries that show signs of higher complexity, such as multiple constraints or explicit references to email chains. This adaptive, multi-strategy approach significantly enhanced the chatbot's retrieval and response performance. For an illustrative example of the chatbot system workflow, see Appendix A.

**4.3 RAG Evaluation (RAG-Eval) Framework for Chatbot Assessment**

To evaluate the effectiveness and reliability of the RAG chatbot, a dedicated LLM-based evaluation framework, RAG-Eval, was developed, inspired by the G-EVAL architecture [35] for a reference-free evaluator for tasks like text summarisation (as well as dialogue generation). RAG-Eval operates by considering all three components of the RAG pipeline—the user query, the retrieved context, and the generated response, enabling a more comprehensive and context-aware evaluation. While the current implementation primarily leverages OpenAI's GPT-4o as the main evaluator for this project, the framework is inherently model-agnostic. RAG-Eval supports a variety of state-of-the-art LLMs and can easily be utilised with models such as Llama, Gemini, Mistral, and others, depending on specific project requirements and the model's capabilities. RAG-Eval systematically assesses chatbot responses against a set of predefined criteria, delivering a comprehensive analysis across key dimensions that underpin user satisfaction and trust.

**4.3.1 Evaluation Metrics**

RAG-Eval evaluates chatbot responses using five key metrics:

1. **Query Relevance:** Measures how effectively the generated response addresses the user's query. A relevant response directly answers the question without including unrelated or redundant information. Here is the definition and step prompts:

    ***Query Relevance Criteria:***

    *Query Relevance (1-5): Measures how well the generated response addresses the user's query.*

    *A relevant response answers the query directly and avoids unrelated or redundant information.*

    ***Query Relevance Steps:***

    *1. Read the user query, source document, and generated response.*

    *2. Identify the main intent of the user query.*

    *3. Assess whether the generated response adequately addresses the user's question.*

    *4. Assign a relevance score from 1 to 5.*

2. **Factual Accuracy:** Assesses the correctness of the information provided, ensuring all facts are accurate and supported by the source document. This metric helps detect and penalise ungrounded or unsupported claims. Here is the definition and steps:

    ***Factual Accuracy Criteria:***

*Factual Accuracy (1-5): Determines if the response is factually accurate, without any errors, based on the content in the source document.*

***Factual Accuracy Steps:***

*1. Compare the generated response with the source document.*

*2. Identify any factual discrepancies, incorrect details, or hallucinations.*

*3. Assign an accuracy score between 1 and 5 based on how well the response aligns with the source facts.*

*4. Review* and *penalise if the response contains any facts or references not found in the source document.*

3. **Coverage:** Evaluates the extent to which the response includes all pertinent information from the source document relevant to the query. A comprehensive response captures all essential points, providing a complete answer. Here is the definition and steps:

   ***Coverage Criteria:***

   *Coverage (1-5): Evaluate whether the response captures all the key points from the source document that are relevant to the query.*

   ***Coverage Steps:***

   *1. Read the user query and the source document.*

   *2. Check if the generated response covers all essential points from the source that answer the query.*

   *3. Assign a coverage score from 1 to 5.*

   *4. Review* and *penalise if important details from the source are missing.*

4. **Coherence:** Examines the logical flow and structure of the response. Information should be well-organised, with ideas presented in a manner that is easy to follow and understand. Here is the definition and steps:

   ***Coherence Score Criteria:***

   *Coherence (1-5): The response should be well-organised and logically structured, with the ideas flowing naturally.*

   *The integration of information from the source should make sense as a cohesive answer to the query.*

   ***Coherence Score Steps:***

   *1. Read the response carefully and assess how well the ideas are structured.*

   *2. Check whether the information from the source document has been combined into a coherent response.*

   *3. Assign a coherence score from 1 to 5.*

5. **Fluency:** Assesses the grammatical correctness and readability of the response. Language should be clear, concise, and free of grammatical errors. Here is the definition and steps:

   ***Fluency Score Criteria:***

   *Fluency (1-5): Evaluates the quality of the language used, including grammar, punctuation, and sentence structure.*

   ***Fluency Score Steps:***

1. Read the response and evaluate its readability, grammatical correctness, and fluency.
2. Assign a fluency score from 1 to 5.

These metrics capture both the content quality and presentation of the chatbot's generated responses, ensuring a holistic evaluation aligned with user expectations for accuracy, completeness, and clarity. It is also worth noting that for both factual accuracy and coverage evaluation, penalisation is applied after the initial scoring. This ensures that any unsupported claims or missed essential facts, identified on a second review, result in an appropriate score adjustment.

**4.3.2 Scoring Methodology**

Each metric is scored on a scale from 1 to 5, as shown in Table 1:

Table 1 Scoring Scale for RAG-Eval Metrics

| Score | Descriptor | Description |
|---|---|---|
| 1 | Poor | Significant issues: criteria and steps not met |
| 2 | Fair | Partial alignment with criteria and steps |
| 3 | Good | Mostly accurate, with some minor issues in alignment with criteria/steps |
| 4 | Very Good | High alignment with criteria/steps, with only minor inconsistencies |
| 5 | Excellent | Flawless performance; all criteria and steps fully met |

To aggregate these scores as a confidence score, each metric has been normalised to a 0 to 1 scale using:

**Normalised Score** = $\frac{Score}{Maximum\ Score}$

**Metric Weights:** Before the aggregation score is computed, weights are assigned to each metric to reflect their importance. In this system, users can specify custom weights for each metric (Query Relevance, Factual Accuracy, Coverage, Coherence, Fluency) according to project needs. Weights must be a list of five values summing to 1.0, ensuring a flexible but consistent influence of each metric. If not specified, default weights are used.

**Weighted Average Calculation:** The confidence score is calculated as a weighted average of the normalised metric scores:

$$Confidence\ Score = \sum_{i=1}^{N} (Normalised\ Score_i \times Metric\ Weight_i)$$

*Where N = 5 (the number of metrics).*

**Default Weights:** By default, greater emphasis is placed on the most critical metrics for retrieval-augmented generation (RAG) applications, such as chatbots. To reflect this, the following weights are used (Table 2):

Table 2: Default Metric Weights for Confidence Score Calculation

| Metric | Default Weight |
|---|---|
| Query Relevance | 0.25 |
| Factual Accuracy | 0.25 |
| Coverage | 0.25 |
| Coherence | 0.125 |
| Fluency | 0.125 |

*Note: Users may override these default values, provided the sum equals 1.0.*

This weighting scheme emphasises the importance of query relevance, factual accuracy, and coverage, which are critical for user trust and satisfaction in information retrieval systems. To enable users to effectively assess the chatbot's performance, the confidence score is presented as a percentage (%).

**Confidence Score (%) = Confidence Score x 100%**

**4.3.3 Implementation Details**

RAG-Eval is implemented by employing prompt engineering (the process of designing and refining input instructions (prompts) to elicit desired responses from LLM to guide the model in solving a particular task). These prompts systematically guide LLM to evaluate each response along the five metrics through carefully structured instructions. The evaluation process involves the following steps:

1. **Evaluation Prompt Design:** For each metric, specific prompts have been crafted that include Evaluation Criteria that clearly define the outlining expectations for each metric and Evaluation Steps that define Step-by-step instructions for the model to follow when assessing the response. This structured approach ensures consistency and objectivity in evaluations.
2. **Automated Evaluation Pipeline:** The evaluation pipeline operates with four activities. The Input takes the user's query, the source document (retriever), the generated response, and the evaluation form, as seen in the description below.
   *Evaluation Prompt Template:*

   *You will be given a user query, a response generated by a language model, and the knowledge source document that was used to generate the response. Your task is to evaluate the response on one specific metric.*

   *Evaluation Criteria: {criteria} = > Evaluation Steps: {steps}*

   *Example: User Query: {query}, Source Document: {document}, Generated Response: {response}, Evaluation Form (scores ONLY): {metric name}*

   The Metric Evaluation (evaluation form) gets the LLM assigned score based on the criteria and steps provided for each metric. The Score Normalisation and Aggregation provide normalised and weighted raw scores to calculate the overall confidence score. Finally, the Output activity generates the individual metric scores and the overall score (confidence score), offering a detailed assessment of response quality.
3. **Integration with Chatbot Interface:** The individual metric scores and the confidence score are integrated into the chatbot system, providing users with immediate feedback on the reliability of responses. This transparency enhances user trust and enables informed decision-making based on the presented information.
4. **Calibration and Validation:** The calibration phase started by selecting a representative set of user queries and corresponding retrieved documents, then asking human reviewers to rate the generated responses along the same five RAG-Eval metrics (query relevance, factual accuracy, coverage, coherence, and fluency). Human-assigned ratings were compared to the automated scores produced by LLM under the initial prompt settings and weighting scheme. Discrepancies guided iterative adjustments in two key areas: Evaluation Prompts refined the wording, order, and emphasis within the LLM prompts to resolve ambiguities that caused the model to overvalue or undervalue specific aspects (like minor factual inconsistencies or borderline coverage cases). Metric Weightings systematically rebalanced weights (especially between coverage, factual accuracy, and query relevance) based on how strongly each factor influenced human perceptions of a "good" response.

Repeated this process until LLM's automated ratings aligned closely with human judgments, ensuring that RAG-Eval scores more reliably reflected human expectations. This iterative calibration loop both improves the metric's consistency and instils greater confidence in its ability to capture response quality across complex, domain-specific queries.

**4.4 Metadata Integration for Enhanced Response Transparency and Verification**

To improve the transparency and reliability of chatbot responses, integrating metadata within generated summaries was experimented with. This allows users to verify information directly against source data, fostering strong user confidence by enabling cross-referencing of essential metadata such as email IDs, conversation thread IDs, timestamps, etc.

**Approach and Setup:** Two strategies for metadata inclusion have been set up.

1. **Fine-Tuning of Large Language Model**: In this first method, the GPT-3.5-turbo Model was fine-tuned—the process of adapting a pre-trained language model to a specific task by further training it on a specialised dataset. A manually crafted synthetic dataset was used for the fine-tuning. This synthetic dataset comprised structured question-answer pairs, where each answer embedded relevant metadata. These fine-tuning enabled responses to align with the RAG model setup, ensuring consistent inclusion of metadata like email IDs and timestamps.

2. **Prompt Engineering:** In this method, the prompt engineering utilised prompt-based instructions to guide GPT-4o in integrating metadata into responses. This has been chosen because of GPT-4 series (including variants like GPT-4.1, which was also used during later testing) shows superior ability to dynamically generate Cypher queries and retrieve specific metadata from the Neo4j knowledge graph. In this prompt engineering, carefully crafted prompts allowed the LLM-based agent to embed metadata while maintaining high contextual accuracy.

While both methods were effective for this task, we opted for GPT-4o without fine-tuning and streamlined the system architecture by using a single model for all chatbot functions. We made this choice in the spirit of the project: retrieval and prompt engineering over finetuning. This also eliminated the need for maintaining additional fine-tuned models and simplified the overall setup. However, developers retain the flexibility to choose the approach that best aligns with their model capabilities and project requirements, whether that involves GPT-4o, GPT-4.1, or any other advanced model.

**4.5 Reproducibility and Open-Source Code**

To ensure reproducibility while protecting proprietary data, all figures in the appendix section are generated using a synthetic dataset created with an LLM (GPT-4). The original dataset used in this project is proprietary and cannot be shared publicly in any form. Therefore, a synthetic dataset with a similar structure is provided for testing and demonstration purposes. The complete project, including the synthetic data, Docker setup, and instructions for local testing, is available in the open-source repository: https://github.com/OlaAkindele/rag_chatbot. The RAG Evaluation (RAG-EVAL) framework is also available as a standalone package at: https://github.com/OlaAkindele/rag_evaluation.

**5. Experimental Results and Analysis**

To assess the performance of the presented chatbot system, empirical experiments were conducted using a diverse set of questions (queries). For each query, three key components: the user query itself, the document retrieved (retrieved backend), and the chatbot's generated response (retrieved frontend), were recorded. These were presented as input into the RAG-Eval framework, allowing for a tripartite (three-input) assessment that directly measures RAG system output.

**5.1 Summarisation Quality Analysis**

The RAG-Eval scores have been compared with established evaluation metrics such as BERTScore and G-EVAL to assess the summarisation quality of the chatbot's responses. While RAG-Eval utilises three inputs, Query, Retriever, and Response to assess the summarisation, the BERTScore and G-EVAL are designed for two inputs. for instance, source text and summarisation text. Because BERTScore and G-EVAL traditionally work by comparing a source text to a summarised text, therefore, treating content returned from the knowledge graph (the retriever output) as the 'source' and the chatbot's generated response as the 'summarisation.' This approach ensures that BERTScore and G-EVAL could still evaluate the results in a standard two-input format (source vs. summary), even though RAG-Eval intrinsically incorporates an additional parameter (the user's query) into its evaluation process.

Table 3 Comparison of Evaluation Metrics for Summarisation Quality

| Metric | Sub-Metric | Conversation (Query, Retriever, Response) | | | | |
|---|---|---|---|---|---|---|
| | | One | Two | Three | Four | Five |
| **BERTScore** (Retriever + Response) | **Precision** | 0.87 | 0.89 | 0.87 | 0.88 | 0.87 |
| | **Recall** | 0.79 | 0.82 | 0.78 | 0.79 | 0.80 |
| | **F1 Measure** | 0.83 | 0.85 | 0.82 | 0.83 | 0.83 |
| **G-EVAL** (Retriever + Response) | **Coherence** | 1.00 | 1.00 | 1.00 | 1.00 | 1.00 |
| | **Consistency** | 1.00 | 1.00 | 1.00 | 1.00 | 1.00 |
| | **Fluency** | 1.00 | 1.00 | 1.00 | 1.00 | 1.00 |
| | **Relevance** | 1.00 | 1.00 | 1.00 | 1.00 | 1.00 |
| **RAG-Eval** (Query + Retriever + Response) | **Query Relevance** | 1.00 | 1.00 | 1.00 | 1.00 | 1.00 |
| | **Factual Accuracy** | 1.00 | 1.00 | 1.00 | 1.00 | 1.00 |
| | **Coverage** | 1.00 | 1.00 | 1.00 | 1.00 | 0.80 |
| | **Coherence** | 1.00 | 1.00 | 1.00 | 1.00 | 1.00 |
| | **Fluency** | 1.00 | 1.00 | 1.00 | 1.00 | 1.00 |
| | **Overall Accuracy @RAG-Eval** | 1.00 | 1.00 | 1.00 | 1.00 | 0.95 |

Although Table 3 shows that embedding-based metrics like BERTScore can quantify token-level similarity between generated responses and source, while G-EVAL, when applied only to (retriever output + generated response) scores coherence, consistency, fluency, and relevance to the source document, though, the both were not designed to explicitly measure "Does this answer address the user's question?". In particular, G-EVAL's evaluation reflects whether the summary captures the major themes of the source document, not whether every piece of information requested in the query (like the metadata of the source document) is present. Overall, all three evaluation methods work. However, RAG-Eval's explicit inclusion of the user's query makes it more sensitive to both factual accuracy and coverage. For example, in Conversations Five, RAG-Eval reports coverage scores of 0.80 alongside factual accuracy of 1.00. This shows that, although the chatbot's response remains factually correct, RAG-Eval penalises even minor omissions or fine-grained facts, such as missing metadata or names (person ID) from lengthy retrieved documents, reflecting its stricter coverage criteria.

### 5.2 Query-Response Mismatch Analysis

In practical applications, the chatbot may sometimes generate responses that are not relevant to the user's actual query due to misinterpretation, errors in the retrieval process, or other unforeseen factors. To simulate and assess this scenario, the queries were deliberately modified while keeping the retrieved documents and chatbot responses the same. For instance, in one scenario, replace the query 'What was Mr. X's discussion on the company's purchase order (PO)?' with 'What was Mr. X's discussion on the company's pull tester training?'. This intentional mismatch evaluates how effectively RAG-Eval and other sub-metrics detect and penalise responses that do not appropriately address the user's query.

Table 4: Evaluation Metrics with Mismatched Queries

| Metric | Sub-metric | Conversation (Query, Retriever, Response) | | | | |
|---|---|---|---|---|---|---|
| | | One | Two | Three | Four | Five |
| **RAG-Eval** (Query + Retriever + Response) | **Query Relevance** | 0.20 | 0.20 | 0.20 | 0.20 | 0.20 |
| | **Factual Accuracy** | 1.00 | 1.00 | 1.00 | 0.80 | 1.00 |
| | **Coverage** | 0.60 | 0.20 | 0.60 | 0.60 | 0.60 |
| | **Coherence** | 0.80 | 0.80 | 0.80 | 0.80 | 0.80 |
| | **Fluency** | 0.80 | 1.00 | 1.00 | 1.00 | 1.00 |
| | Overall Accuracy @RAG-Eval | 0.65 | 0.58 | 0.68 | 0.63 | 0.68 |

The significant decrease in Query Relevance and Overall Accuracy illustrates how changing only the question, even though the retriever outputs and chatbot responses remain unchanged, leads to misalignment that RAG-Eval readily identifies. Notice that if we had asked G-EVAL (which is designed with two inputs) to score these same pairs, it would have no direct way to see that the user's question has changed. By explicitly incorporating "(user query + retrieved document + generated response)," RAG-Eval penalises models for not answering the updated question, an evaluation dimension that G-EVAL was not originally built to handle.

### 5.3 Factual Consistency Analysis

To further test how well the chatbot system aligns with factual data (source data) and evaluate the presented model's ability to detect such instances, the original responses were replaced with generic statements related to the query topics, but not factual data from the retrieved documents.

Table 5: Evaluation Metrics for Factual Consistency

| Metric | Sub-metric | Conversation (Query, Retriever, Response) | | |
|---|---|---|---|---|
| | | One | Two | Three |
| **RAG-Eval** (Query + Retriever + Response) | **Query Relevance** | 0.40 | 0.20 | 0.20 |
| | **Factual Accuracy** | 0.20 | 0.20 | 0.20 |
| | **Coverage** | 0.20 | 0.20 | 0.20 |
| | **Coherence** | 0.40 | 0.20 | 0.20 |
| | **Fluency** | 1.00 | 1.00 | 1.00 |
| | Overall Accuracy @RAG-Eval | 0.38 | 0.30 | 0.30 |

The results show a significant decline in Query Relevance, Factual Accuracy, and Coverage scores, indicating that the responses do not adequately address the queries or reflect the retrieved content. Despite relatively high Fluency scores, which suggest well-formed language, RAG Eval effectively identifies these factual misalignments between the query, retrieved document, and response. This demonstrates RAG-Eval's capability to detect fluent but ungrounded responses, which is crucial for developing a reliable RAG chatbot.

All in all, this empirical analysis and experimental evaluation also provided additional insights into interpreting the confidence scores (Overall Accuracy @RAG-Eval) practically: Scores above 74% typically indicate high factual accuracy and strong alignment with source documents, although some minor or fine-grain details may be missing. Scores between 50% and 74% generally suggest that while the response remains grounded in the correct source document, it may not comprehensively or accurately address every aspect of the question. Scores below 50% likely denote significant misalignment or factual inaccuracies. This interpretation helps both developers and users rapidly assess response reliability, making informed decisions about the trustworthiness of chatbot-generated answers.

## 6. Discussion

Having presented the architecture, evaluation framework, and empirical results for the RAG chatbot system, this section critically examines the broader implications and practical significance of the work. It discusses the key findings and contributions, evaluates how the combined use of a knowledge graph and RAG-Eval advances scalable, transparent chatbot solutions, and reflects on the system's real-world strengths and shortcomings.

### 6.1 Key Findings and Contributions

This work confirms several important contributions to the development and evaluation of large-scale, domain-specific chatbots:

- **Leverage of Structured Data and Knowledge Graphs for Retrieval:**

  Integrating a knowledge graph with vector embeddings preserves both context and relationships, enabling precise and accurate information retrieval. Unlike methods that convert structured data into an unstructured text file [49], this approach maintains semantic integrity, mitigates ambiguity, and allows for reliable answers to project-specific queries.

- **RAG-Eval's Multi-Faceted, User-Facing Evaluation:**

  By evaluating each response using the user query, retrieved data, and generated output, RAG-Eval captures critical quality dimensions—query relevance, factual accuracy, coverage, fluency, and coherence. This approach reliably identifies responses that may be fluent yet incomplete or off-topic, which more traditional NLP metrics (such as ROUGE or BERTScore) might overlook.

- **Metadata-Enriched Responses and Source Verification for Transparency:**

  By embedding source references (email IDs, revision IDs, timestamps) into chatbot replies and presenting a confidence score, the system enables immediate verifiability, efficient fact-checking, and user trust, particularly in high-demand environments where factual accuracy is critical.

- **Scalability and Handling of Large Datasets:**

  The chatbot system efficiently processes a substantial dataset comprising 108,000 rows and totalling approximately 48 million words, demonstrating scalability and performance beyond related works that utilise smaller corpora.

- **Adaptive Handling of Ambiguous or Failed Queries:**

  When a Cypher query yields no results, the chatbot automatically falls back to alternative strategies (such as vector search), iterating until a contextually accurate response emerges. This ensures that even vague or partially formed queries can receive meaningful answers.

Empirical analysis during evaluation suggests that a confidence score of 75% or higher provides users with high trust in the chatbot's responses. Scores below this threshold warrant caution, though the cutoff can be adjusted depending on the specific use case or LLM in use. While RAG-Eval offers a scalable and automated approach for evaluating chatbot responses, it should be seen as complementary to human judgment, not a replacement. The inclusion of reference metadata in outputs also enhances the efficiency of human review when needed.

### 6.2 Limitations and Challenges

Despite the advances, several limitations remain:

- **Database Scalability and Response Latency:**

As the Neo4j AuraDB graph database grows, a slight increase in response latency has been observed, though not to a critical extent. Further scaling may require database upgrades or more efficient storage/retrieval solutions.

- **Dependence on LLMs for Cypher Query Generation:**

  The chatbot relies on an AI agent to generate accurate Cypher queries. Occasionally, suboptimal queries impact retrieval effectiveness and necessitate fallback strategies, such as vector search or further iterations, to ensure satisfactory results.

- **Minor Variability in RAG-Eval Scores:**
  Slight variations in overall accuracy were observed upon re-evaluation (for example, 96% to 97% or 95%) on a few occasions. These discrepancies are minimal and considered negligible for practical purposes.

## 7. CONCLUSION

This research has introduced an advanced Retrieval-Augmented Generation (RAG) chatbot that combines a knowledge graph with vector-based retrieval to manage large, structured datasets. By implementing real-time evaluation with the novel RAG-Eval tripartite framework, which operates independently of the chatbot to assess the user query, retrieved content, and generated response together, the system provides users with immediate and transparent feedback on response quality. This enables quick identification of factual inaccuracies or misalignments without interfering with the chatbot's primary output process. RAG-Eval's confidence score, promptly displayed to the users, acts as an early warning signal; low scores indicate possible factual inaccuracies or misalignment with the query. The chatbot's metadata-enriched summaries further enhance transparency and trust by enabling users to quickly verify sources such as email IDs and timestamps. Experimental results show that RAG-Eval delivers a more comprehensive assessment, capturing domain-specific details, factual accuracy, and query relevance in real time. Although the chatbot system approach occasionally depends on fallback vector search when Cypher query generation fails, the overall framework remains scalable for large datasets with complex relationships. Future work will explore ways to streamline and optimise both Cypher query construction and database performance and extend the chatbot system to unstructured data scenarios. Taken together, this research demonstrates a path toward building more reliable and trustworthy RAG chatbots, where real-time evaluation helps ensure both clarity and factual consistency.

**Data Availability Statement (DAS)**

"Raw data used in this work are not publicly available to preserve individuals' privacy, along with company privacy under the European General Data Protection Regulation."

# APPENDIX

**Appendix A: Example Workflow Demonstration with Simulated Dataset**

To facilitate reproducibility and practical demonstration, a synthetic engineering email dataset accompanies the project's open-source repository. Fig. 4 and 5 illustrate typical backend and frontend interactions, showcasing query handling, response summarisation, and metadata inclusion.

**Backend Interaction Example (Fig. 4)**:

Upon receiving the query *"What are the main contents of the emails titled 'underdeck gantry design discussion'?"*, the backend generates and executes a Cypher query, summarising retrieved emails with essential metadata (email IDs, timestamps).

**Fig. 4** Backend log demonstrating automatic Cypher query generation and summarised response composition.

**Frontend User Interface (Fig. 5)**:

The corresponding frontend displays a structured, metadata-enriched response along with a confidence score (RAG-Eval) clearly indicating response trustworthiness. An additional example demonstrates query handling for "3D Modelling," highlighting the system's response to sender and recipient names.

**Experimental Database Assistant Chatbot**

**You:** What are the main contents of the emails titled 'underdeck gantry design discussion'?

**Bot:** The emails titled "underdeck gantry design discussion" (emailIds: 9806201052, 9140646526, 9366191665, and 2884024936) cover the following main topics: - On 2024-02-03 (emailId: 9806201052), the team was asked to review attached design drafts to finalize the underdeck gantry design. - On 2022-03-02 (emailId: 9140646526), there was a request to ensure the design meets all regulatory standards. - On 2022-08-04 (emailId: 9366191665), a discussion was initiated regarding the structural integrity of the design. - On 2023-10-01 (emailId: 2884024936), the focus was on reviewing the safety and operational efficiency aspects of the gantry design. These emails collectively document an ongoing discussion about finalizing the design, ensuring regulatory compliance, addressing structural integrity, and reviewing safety and operational efficiency for the underdeck gantry. All relevant email IDs and dates are provided for reference.
Confidence score: 100.00%

**You:** I am looking for any emails related to 3D modeling requests. If any are found, I would like to know what they are about and who is involved in the discussion.

**Bot:** There is one email in the database related to 3D modeling requests: - Email ID: 9188944267 - Date Received: 2024-03-20 11:00:02 UTC - Subject: 3D Modeling Requests - Summary: The email discusses that several 3D modeling requests have been received for a new project and asks the team to review the attached request list. - Sender: Person_08ca204f2984 - Receivers: Person_da7eed0ef27f, Person_9af157c7a573, Person_17bfc41198c6 This is the only email specifically mentioning 3D modeling requests. The discussion involves the sender (Person_08ca204f2984) and three recipients (Person_da7eed0ef27f, Person_9af157c7a573, and Person_17bfc41198c6), all referenced by their unique IDs. The email was received on March 20, 2024.
Confidence score: 100.00%

**Fig. 5** Frontend interface displaying structured chatbot responses and confidence scores